\documentclass[twocolumn,tightenlines,superscriptaddress,eqsecnum,floats,aps,prd,showpacs]{revtex4-2}
\usepackage{amssymb}
\usepackage{stmaryrd}
\usepackage{amsmath,amssymb,amsfonts}

\usepackage{xeCJK} 
\usepackage{fontspec} 

\usepackage{graphicx}
\usepackage{subcaption}
\usepackage{color} 
\usepackage{hyperref}
\usepackage{mathrsfs}

\begin{document}

	\title{Periodic orbits and  gravitational waveforms of black holes in bumblebee gravity}
    
	\author{Zijian Shi}
	\affiliation{School of Physics and Optoelectronics, South China University of Technology, Guangzhou 510641, China}
	
	\author{Xiangdong Zhang}
	\thanks{Corresponding author. scxdzhang@scut.edu.cn}
	\affiliation{School of Physics and Optoelectronics, South China University of Technology, Guangzhou 510641, China}
	
	\author{Yunlong Liu}
	\thanks{Corresponding author. phliuyunlong@kust.edu.cn}
	\affiliation{School of Science, Kunming University of Science and Technology, Kunming 650093, China}

    \date{\today}

	\begin{abstract}
    In this paper, we investigate the dynamics of massive particles and the associated gravitational waveforms in the spacetime of a black hole within the framework of Einstein-Bumblebee gravity. Our analysis encompasses both charged and uncharged black hole configurations, with a particular focus on the spontaneous Lorentz symmetry breaking mechanism inherent to this model, which is governed by a dimensionless coupling parameter $l$. We analyze the geodesic equations and the effective potential to determine the allowed parameter space for bound orbits, demonstrating that in the charged case, both the Lorentz-violating parameter $l$ and the electric charge $Q$ significantly enhance the confinement capacity of the potential, thereby broadening the energy and angular momentum windows for bound states. A key focus is placed on the classification and properties of periodic orbits, characterized by rational frequency ratios using the whirl, zoom, and vertex taxonomy. We demonstrate that in the uncharged case ($Q=0$), the radial effective potential and standard innermost stable circular orbit (ISCO) properties are degenerate with those of a Schwarzschild black hole. However, despite this degeneracy in static potential properties, the structure of periodic orbits exhibits qualitative differences, providing a possible observational signature that can break this degeneracy. Finally, we compute the corresponding gravitational waveforms extracted from these periodic orbits using the quadrupole formula. The results reveal that $l$ and $Q$ introduce contrasting phase-shifting effects on the waveforms. This suggests that bumblebee gravity leaves measurable imprints on gravitational-wave signals that could be detected by future space-based gravitational-wave observatories.

	\end{abstract}
	\maketitle

	\section{introduction}
	General relativity (GR) stands as one of the pillars of modern physics, providing an elegant and empirically successful description of gravity. Its most dramatic predictions—the existence of black holes and the propagation of gravitational waves—have transitioned from theoretical curiosities to observational cornerstones, thanks to groundbreaking experiments like LIGO/Virgo \cite{LIGOScientific:2016aoc} and the Event Horizon Telescope (EHT) \cite{EventHorizonTelescope:2019dse, EventHorizonTelescope:2022wkp}. These discoveries have catapulted black holes from mere mathematical solutions to prime astrophysical laboratories, where the dynamics of matter and light in the most extreme gravitational fields can be used to test the very fabric of spacetime \cite{Barack:2018yly}.
    
    Despite its triumphs, GR is famously incompatible with quantum mechanics at a fundamental level, motivating the search for a unified theory. A promising avenue involves exploring potential low-energy relics of quantum gravity. One such relic is spontaneous Lorentz symmetry breaking, a phenomenon where the vacuum state of the universe picks out a preferred direction, thereby violating the isotropic symmetry of spacetime \cite{Gambini:1998it, Ellis:1999uh, Kostelecky:1988zi, Kostelecky:1989jp}. The Einstein-Bumblebee model offers a concrete and well-motivated effective field theory framework to investigate such Lorentz-violating effects \cite{Kostelecky:2003fs}. In this model, Lorentz symmetry is spontaneously broken by a dynamical vector field (the ``bumblebee'' field) that couples non-minimally to the curvature of spacetime, leading to modified gravitational dynamics.
    
    A critical test for any modified gravity theory is its prediction for compact astrophysical objects, particularly black holes. The properties of black hole solutions—their event horizon structure, orbital dynamics, and emitted radiation—carry unique imprints of the underlying gravitational theory. Recently, significant progress has been made within the bumblebee gravity framework \cite{Casana:2017jkc, Maluf:2020kgf, Liu:2024axg, Li:2025rjv, Liu:2025oho, Bailey:2025oun, Xu:2022frb, Marques:2023suh, AraujoFilho:2025rvn, Ovcharenko:2026rvj}. By incorporating a non-minimal coupling between the bumblebee field and an electromagnetic field, Liu \textit{et al.} derived a new class of exact charged black hole solutions \cite{Liu:2024axg}. Among these is a particularly relevant solution: a charged, spherically symmetric black hole without a cosmological constant. This solution represents a Lorentz-violating generalization of the classic Reissner-Nordstr\"om (RN) black hole, characterized by an additional dimensionless parameter, $l$, which quantifies the strength of Lorentz symmetry breaking. This novel spacetime geometry provides a controlled, parametrized deviation from GR, ideal for studying how fundamental symmetry violations might manifest in observable astrophysical phenomena.

    This novel charged bumblebee black hole solution has attracted considerable attention. Subsequent research has explored various physical aspects of this spacetime \cite{Liu:2024axg, Lambiase:2024uzy, Ali:2025znb, Singh:2025hor, Singh:2025hor, Xu:2025jvk, Singh:2025tvk, Li:2025itp, Singh:2025fuv, deOliveira:2025yeo, Li:2026uwx}. The existing literature has addressed the thermodynamic properties \cite{Liu:2024axg, Ali:2025znb}, including Hawking spectra \cite{Singh:2025tvk}. The optical appearance, encompassing phenomena like black hole shadows \cite{Liu:2024axg, Singh:2025hor, Xu:2025jvk, Singh:2025fuv}, photon rings \cite{Xu:2025jvk}, and strong gravitational lensing \cite{Singh:2025tvk}, has been investigated in detail. In particular, studies have used EHT data to constrain the Lorentz-violating parameter $l$ \cite{Lambiase:2024uzy, Xu:2025jvk}. Furthermore, field perturbations have been studied through the calculation of quasinormal modes \cite{Singh:2025hor, Li:2025itp, Singh:2025fuv, deOliveira:2025yeo}, providing insights into the dynamical stability of the black hole.
    
    The study of test particle dynamics around black holes is of paramount importance in the era of gravitational-wave astronomy. A prime example is extreme-mass-ratio inspirals (EMRIs), where a stellar-mass compact object orbits a supermassive black hole---a system that can be modeled as a test particle moving in a curved spacetime \cite{Babak:2017tow}. These systems are recognized as primary sources for future space-based gravitational-wave observatories, including LISA \cite{LISA:2017pwj, LISA:2022yao}, TianQin \cite{TianQin:2015yph, TianQin:2020hid}, Taiji \cite{Hu:2017mde, Ruan:2018tsw}, and DECIGO \cite{Kawamura:2020pcg}. Because the inspiral waveform encodes the precise geometry of the central object, it is highly sensitive to the underlying spacetime structure \cite{Amaro-Seoane:2007osp, Gair:2012nm, Berti:2015itd, Cardenas-Avendano:2024mqp}. Consequently, modifications to the geometry, such as that introduced by the Lorentz violation, can leave characteristic imprints on both the orbital dynamics and the resulting gravitational radiation. Therefore, geodesic analysis is not merely a theoretical exercise, but a necessity for constructing accurate waveform templates for realistic astrophysical scenarios and various theories of gravity \cite{Liu:2024qci, Liu:2024lda,Du:2025}. Notably, recent studies have examined time-like geodesics in this charged bumblebee black hole spacetime \cite{Xu:2025jvk, Li:2026uwx}, specifically analyzing the circular geodesics and innermost stable circular orbit (ISCO).

    Despite these advancements, a critical class of motion remains largely unexplored in this spacetime: periodic orbits \cite{Levin:2008mq}. In the relativistic regime, generic bound orbits around black holes are not closed; instead, they typically precess. A periodic orbit occurs when the radial and angular oscillation frequencies form a rational ratio, causing the particle to retrace its path after a finite number of cycles. These orbits can be characterized by integers $(w,z,v)$ representing whirl, zoom, and vertex numbers, creating a ``periodic table'' of orbital structures. Other types of bound orbits can often be viewed as perturbations of these fundamental periodic orbits. While such periodic orbits have been extensively investigated in various black hole spacetimes \cite{Misra:2010pu, Haroon:2025rzx, Zhang:2025wni, Yang:2024lmj, Gong:2025mne, Chen:2025aqh, Babar:2017gsg, Liu:2018vea, Wei:2019zdf, Deng:2020yfm, Zhou:2020zys, Yao:2023ziq, Tu:2023xab, Junior:2024tmi, QiQi:2024dwc, Bravo-Gaete:2026com}, their properties in the context of bumblebee gravity, particularly for the charged bumblebee black hole, remain unexplored. In this paper, we bridge this gap by conducting a detailed analysis of periodic orbits and their associated gravitational waveforms in this spacetime.

    The remainder of this paper is organized as follows. In Sec.~\ref{section2}, we briefly review the spacetime geometry of the static, spherically symmetric charged black hole within Einstein-Bumblebee gravity. In Sec.~\ref{section3}, we derive the geodesic equations for a test particle, analyze the characteristics of the radial effective potential, and determine the allowed parameter space for bound orbits, with particular attention paid to the uncharged ($Q=0$) limit. Sec.~\ref{section4} is dedicated to the study of periodic orbits, where we employ the whirl, zoom, and vertex taxonomy to classify the trajectories and systematically investigate the dependence of the rational frequency ratio on various spacetime and orbital parameters. In Sec.~\ref{section5}, we utilize the quadrupole formula to compute the gravitational waveforms extracted from these periodic orbits, highlighting the distinct dephasing signatures introduced by the Lorentz-violating parameter and the electric charge. Finally, we summarize our main findings and conclude the paper in Sec.~\ref{section6}.

    \section{THE CHARGED BUMBLEBEE BLACK HOLE}\label{section2}

    In this section, we review the construction of the static, spherically symmetric charged black hole solution within the framework of Einstein-Bumblebee gravity \cite{Liu:2024axg}. The theory extends GR by introducing a vector field $B_\mu$, known as the bumblebee field, which non-minimally couples to the spacetime curvature and triggers spontaneous Lorentz symmetry breaking. 
    
    The dynamics of the system are governed by the action: 
    \begin{equation}
    \begin{split}
    S = & \int d^4x \sqrt{-g} \left[ \frac{1}{2\kappa} (R + \xi B^\mu B^\nu R_{\mu\nu}) \right. \\
    & \left. - \frac{1}{4} B_{\mu\nu}B^{\mu\nu} - V(B^\mu B_\mu \pm b^2) \right] + S_M,
    \end{split}
    \end{equation}
    where $\kappa = 8\pi G$ is the gravitational coupling constant, $\xi$ is the non-minimal coupling constant between gravity and the bumblebee field. $B_\mu$ is the bumblebee vector field with its strength tensor $B_{\mu\nu} = \partial_\mu B_\nu - \partial_\nu B_\mu$, which is similar to the electromagnetic field. Spontaneous Lorentz symmetry breaking is driven by the potential $V$, which ensures that the bumblebee field acquires a non-zero vacuum expectation value (VEV). The potential $V$ is designed to have a minimum at $B^\mu B_\mu \pm b^2 = 0$, where $b$ is a positive constant. At this minimum, the field $B_\mu$ develops a VEV $\langle B_\mu \rangle = b_\mu$, satisfying the constraint $b^\mu b_\mu = \mp b^2$. To incorporate the electromagnetic sector, we consider a matter action $S_M = \int d^4x \sqrt{-g} \mathcal{L}_M$, where the electromagnetic field $A_\mu$ is non-minimally coupled to the bumblebee vector field: 
    \begin{equation}
    \mathcal{L}_M = \frac{1}{2\kappa} \left( F^{\mu\nu}F_{\mu\nu} + \gamma B^\mu B_\mu F^{\alpha\beta} F_{\alpha\beta} \right).
    \end{equation}
    Here, $F_{\mu\nu} = \partial_\mu A_\nu - \partial_\nu A_\mu$ is the Maxwell tensor and $\gamma$ represents the coupling coefficient between the two fields. We seek a static, spherically symmetric metric of the form: 
    \begin{equation}
    ds^2 = -A(r)dt^2 + S(r)dr^2 + r^2(d\theta^2 + \sin^2\theta d\phi^2).
    \end{equation}
    The bumblebee field is assumed to be fixed at its VEV, $B_\mu = b_\mu$. Following previous work \cite{Casana:2017jkc}, we choose a purely radial, spacelike VEV configuration: 
    \begin{equation}
    b_\mu = \left(0, b_r(r), 0, 0\right).
    \end{equation}
    The condition that this VEV has a constant norm, $b_\mu b^\mu = b^2$, directly determines its functional form as: 
    \begin{equation}
    b_r(r) = b \sqrt{S(r)}.
    \end{equation}
    For the electromagnetic sector, we assume a purely electric potential $A_\mu = (\Phi(r), 0, 0, 0)$. To derive the specific charged solution, we impose the vacuum conditions $V = 0$ and $V' = 0$ (which can be realized by a quadratic potential $V = \frac{\lambda}{2} (B^\mu B_\mu - b^2)^2$). Furthermore, to ensure the tractability of the resulting field equations, the coefficient is specifically chosen as $\gamma = \frac{\xi}{2+l}$, where $l = \xi b^2$ is a dimensionless parameter characterizing the strength of the Lorentz violation. By varying the action with respect to the metric $g_{\mu\nu}$, the bumblebee field $B_\mu$, and the gauge field $A_\mu$, one obtains a set of coupled field equations. By solving these equations under the aforementioned constraints, one arrives at the following charged spherically symmetric solution \cite{Liu:2024axg}: 
    \begin{equation}
    A(r) = 1 - \frac{2M}{r} + \frac{2(1+l)Q^2}{(2+l)r^2},
    \end{equation}
    \begin{equation}
    S(r) = \frac{1+l}{A(r)},
    \end{equation}
    \begin{equation}
    \Phi(r) = \frac{Q}{r},
    \end{equation}
    where $Q$ is an integration constant related to the electric charge. This solution represents a RN-like black hole within the framework of bumblebee gravity. In the limit $l \to 0$, it smoothly reduces to the standard RN solution of general relativity. The Lorentz-violating parameter $l$ manifests as a modification to the coefficient of the $Q^2/r^2$ term and introduces a global factor $(1+l)$ in the $g_{rr}$ component. While $A(r) \to 1$ as $r \to \infty$, we have $S(r) \to 1+l$, indicating that the spacetime is not asymptotically Minkowskian due to the presence of the Lorentz-violating background field. The solution exhibits two horizons and its causal structure, represented by the Penrose diagram, is identical to that of the RN black hole.

	\section{GEODESICS, EFFECTIVE POTENTIAL, AND BOUND ORBITS}\label{section3}

    \subsection{Geodesic}\label{Geodesic}
    
    Consider the dynamics of a massive test particle in the charged bumblebee black hole. The particle's motion can be described by the Lagrangian \cite{Chandrasekhar:1985kt}:
    \begin{equation}\label{Lagrangian}
        \mathscr{L} = \frac{m}{2}g_{\mu\nu}\frac{dx^\mu}{d\tau}\frac{dx^\nu}{d\tau} ,
    \end{equation}
    where $\tau$ represents the proper time and $m$ is the particle's mass. The corresponding generalized momentum is obtained as:
    \begin{equation}
        p_\mu = \frac{\partial \mathscr{L}}{\partial \dot{x}^\mu} = g_{\mu\nu}\dot{x}^\nu ,
    \end{equation}
    with the overdot denoting differentiation with respect to $\tau$. By combining the metric of the black hole with the momentum equation, we find the full set of equations describing the particle's motion:
    \begin{align}
     p_t &= -m \dot{t} \left(1 - \frac{2M}{r} + \frac{2(1+l)Q^2}{(2+l)r^2}\right) = -E ,\\
     p_\phi &= m r^2 \sin^2\theta \dot{\phi} = L ,\\
     p_r &= \frac{(1+l)m\dot{r}}{1 - \frac{2M}{r} + \frac{2(1+l)Q^2}{(2+l)r^2}} ,\\
     p_\theta &= m r^2 \dot{\theta},
   \end{align}
    where $E$ and $L$ denote, respectively, the energy and orbital angular momentum of the particle. Using these conserved quantities, the velocity components $\dot{t}$ and $\dot{\phi}$ can be expressed as:
    \begin{align}
        \dot{t} &= \frac{E}{m\left(1 - \frac{2M}{r} + \frac{2(1+l)Q^2}{(2+l)r^2}\right)} ,\\
        \dot{\phi} &= \frac{L}{m r^2 \sin^2\theta}.
    \end{align}
    We can simplify the problem by focusing on motion in the equatorial plane. This means we set $\theta = \pi/2$ and consequently $\dot{\theta} = 0$. Additionally, the four-velocity of a massive particle must satisfy the normalization condition $g_{\mu\nu} \dot{x}^\mu \dot{x}^\nu = -1$. 
    With these conditions, the original Lagrangian \eqref{Lagrangian} reduces to:
    \begin{equation}
        \label{Lr}
        \frac{L^2}{2 m r^2} + \frac{(1 + l) m^2 \dot{r}^2 - E^2}{2 m \left(1 - \frac{2M}{r} + \frac{2(1+l)Q^2}{(2+l)r^2}\right)} = -\frac{m}{2} .
    \end{equation}

    In the following, we adopt normalized parameters as follows: 
    \begin{align}
        \bar{E} = \frac{E}{m}; \quad \bar{L} =  \frac{L}{m M}; \quad \bar{Q} = \frac{Q}{M}; \quad \bar{r}= \frac{r}{M}; \quad \bar{t}=\frac{t}{M},
    \end{align}
    and we shall omit the bars on the normalized parameters. For example, in the following text, $E$ actually means $\bar{E}$.
    
    Finally,  Eq.\eqref{Lr} can be rearranged into a standard form for radial motion:
    \begin{equation}\label{rdotsquare}
        \dot{r}^2 = \frac{1}{1+l} \left( E^2 - V_{\text{eff}}^2 \right) ,
    \end{equation}
    where the effective potential $V_{\text{eff}}$ is defined as
    \begin{equation}\label{Veff}
        V_{\text{eff}} = \sqrt{\left(1 + \frac{L^2}{r^2}\right) \left(1 - \frac{2}{r} + \frac{2(1+l)Q^2}{(2+l)r^2}\right)} .
    \end{equation}

	\subsection{Effective potential}\label{Effective_potential}
\begin{figure*}[!htb]
    \centering
    \subfloat[Fixed $L =3.5$ and $Q=0.2$]{
        \includegraphics[width=0.32\textwidth]{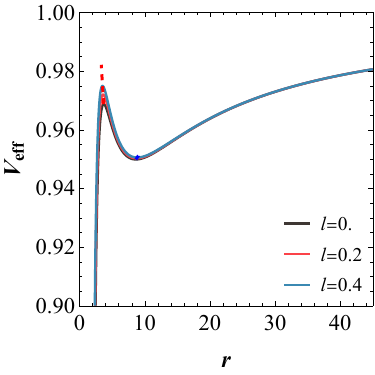}
        \label{Veff_r_l}
    }
    \subfloat[Fixed $L =3.5$ and $l=0.2$]{
        \includegraphics[width=0.32\textwidth]{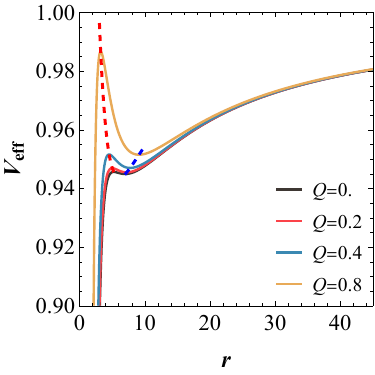}
        \label{Veff_r_Q}
    }
    \subfloat[Fixed $Q=0.7$ and $l = 0.2$]{
        \includegraphics[width=0.32\textwidth]{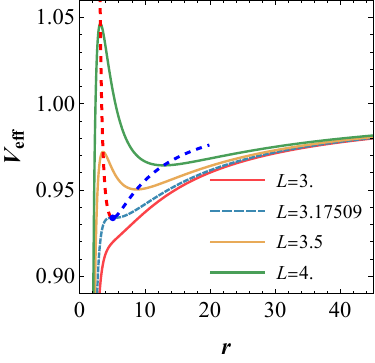}
        \label{Veff_r_L}
    }
    \caption{The radial effective potential $V_{\text{eff}}(r)$ for massive particles. In each panel, the red and blue dashed lines trace the loci of the potential maxima and minima, respectively.}
    \label{Veff_r}
    \end{figure*}

	To investigate the radial motion of the test particles, we analyze the behavior of the effective potential. The radial component of the motion is governed by Eq.~\eqref{rdotsquare}. The physically allowed regions for motion are defined by the condition $E \ge V_{\text{eff}}(r)$. The extrema of $V_{\text{eff}}$ characterize the landscape of the radial motion. While a local minimum corresponds to a stable circular orbit and marks the bottom of a potential well, a local maximum represents an unstable circular orbit and acts as a potential barrier. The interplay between these extrema defines the existence and range of bound orbits, where particles are trapped within the potential well, oscillating between two radial turning points. In Fig.~\ref{Veff_r}, we present the behavior of the effective potential $V_{\text{eff}}(r)$ for different sets of parameters to illustrate how the Lorentz-violating parameter $l$, the charge $Q$, and the angular momentum $L$ modify the orbital landscape.

    In Fig.~\ref{Veff_r_l}, we investigate the influence of the Lorentz-violating parameter $l$ with fixed angular momentum $L=3.5$ and charge $Q=0.2$. As $l$ increases from $0$ to $0.4$, the entire profile of the effective potential is shifted upward. Consequently, the energy levels of both the unstable circular orbit (potential barrier) and the stable circular orbit (potential well) increase as $l$ grows. This indicates that the bumblebee background field exerts an ``uplifting'' effect on the orbital landscape, effectively increasing the energy required for a particle to maintain a bound orbit or to overcome the centrifugal barrier.
    
    A qualitatively similar behavior is observed in Fig.~\ref{Veff_r_Q}, where the electric charge $Q$ is varied. Much like the Lorentz-violating parameter $l$, an increase in $Q$ leads to a systematic elevation of the potential profile. Notably, for increasing $l$ or $Q$, the height of the potential barrier rises more prominently than the bottom of the potential well. This differential shift effectively expands the energy window that allows for bound states, suggesting that these parameters enhance the confinement capacity of the potential landscape.
    
    Finally, we vary the angular momentum $L$ in Fig.~\ref{Veff_r_L}. The red and blue dashed lines track the evolution of the unstable and stable circular orbits as $L$ changes. As $L$ decreases, these two extrema approach each other, narrowing the radial range where bound orbits are permitted. At a critical value $L_c$ (approximately $L_c \approx 3.17509$ for the given parameters), the maximum and minimum coalesce at an inflection point (represented by the blue dot in the figure). This point corresponds to the ISCO. For $L < L_c$, the potential well disappears entirely and the effective potential becomes monotonic, implying that no stable bound orbits can exist and particles will either plunge into the black hole or be scattered to infinity, depending on their energy.

	\subsection{The allowed parameter space for the bound orbits}\label{phase_space}

\begin{figure*}[!htb]
	\centering
    \subfloat[Fixed $l=0.2$]{
        \includegraphics[width=0.45\textwidth]{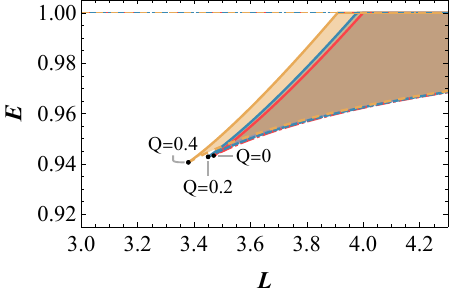}
        \label{EL_Q}
    }
    \subfloat[Fixed $Q=0.2$]{
        \includegraphics[width=0.45\textwidth]{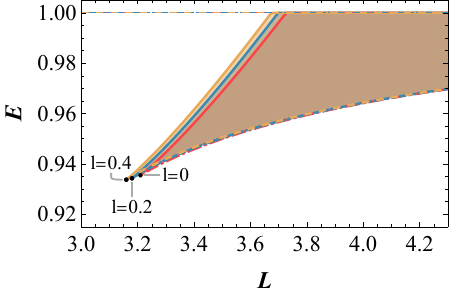}
        \label{EL_l}
    }
		\caption{  The allowed parameter space $(E, L)$ for bound orbits with different $Q$ or $l$. The shaded regions indicate the energy-angular momentum combinations that permit bound motion. The black dots at the leftmost tip of each region denote the ISCO points.  }
		\label{EL}
	\end{figure*}

    In this subsection, we investigate the range of energy $E$ and angular momentum $L$ that permit the existence of bound orbits. As illustrated in Fig.~\ref{EL}, for a given angular momentum $L$, a particle can only be in a bound state if its energy falls within a specific interval, defined by the extrema of the effective potential.

    Fig.~\ref{EL_Q} illustrates the effect of the electric charge $Q$ on the allowed parameter space. As $Q$ increases, the entire allowed region shifts toward the lower-left direction. Specifically, the ISCO point (the black dot in the figure) moves toward lower values of $L$ and lower values of $E$. This suggests that a larger charge $Q$ enables particles with lower angular momentum to maintain bound orbits at a lower energy cost.
    
    A similar trend is observed for the Lorentz-violating parameter $l$ in Fig.~\ref{EL_l}. With the increase of $l$, the permitted region for bound orbits also shifts toward smaller $L$ and lower $E$. The expansion of the shaded area with increasing $l$ further confirms that Lorentz violation strengthens the trapping effect of the black hole's gravitational field. In summary, both the electric charge and the Lorentz-violating parameter contribute to a broader parameter space for bound orbits, reflecting the modified spacetime geometry in bumblebee gravity.

    \subsection{The special case of vanishing charge: $Q=0$ limit}\label{Q0}
    We are especially interested in the limit where the black hole charge vanishes, $Q=0$. From the effective potential \eqref{Veff}, it is evident that when $Q=0$, the potential reduces to:
    \begin{equation}\label{VeffQ0}
        V_{\text{eff}} |_{Q=0} = \sqrt{\left(1 + \frac{L^2}{r^2}\right) \left(1 - \frac{2}{r}\right)}.
    \end{equation}
    In this limit, the Lorentz-violating parameter $l$ is absent from the effective potential. This leads to several key observations regarding the orbital properties.
    
    First, the landscape of the radial effective potential becomes independent of the Lorentz-violating background. Consequently, the bifurcation of the $V_{\text{eff}}$ curves for different values of $l$, as observed in Fig.~\ref{Veff_r_l} for the charged case, would entirely vanish when $Q=0$. In this limit, the potential profiles for $l=0$, $l=0.2$, and $l=0.4$ would perfectly overlap, rendering the uplifting effect of the bumblebee field on the potential invisible.
    
    Second, the allowed parameter space $(E, L)$ for bound orbits, discussed in Sec.~\ref{phase_space}, would exhibit a similar degeneracy. Since the extrema of $V_{\text{eff}}$ (which define the boundaries of the shaded regions in Fig.~\ref{EL}) are determined solely by the Schwarzschild-like potential \eqref{VeffQ0} when $Q=0$, the shaded regions in Fig.~\ref{EL_l} for different $l$ would become identical and perfectly superimposed. The properties of ISCO, such as the radius, energy, and angular momentum, would thus be indistinguishable from those in standard GR.
    
    However, it is crucial to emphasize that the independence of $V_{\text{eff}}$ from $l$ does not imply that the particle motion are unaffected by Lorentz violation. The radial velocity $\dot{r}$ governed by Eq.~\eqref{rdotsquare} still retains the scaling factor $1/(1+l)$. This indicates that even if the effective potential is identical to that of a Schwarzschild black hole, the rate at which a particle traverses the potential—and consequently the accumulation of the azimuthal angle $\Delta \phi$—is significantly modified by $l$. This subtle distinction between the degenerate potential landscape and the non-degenerate physical motion provides a unique signature of pure bumblebee gravity, which we shall analyze in the context of periodic orbits and frequency ratios in the subsequent section.

	\section{PERIODIC ORBITS}\label{section4}
	Along a periodic orbit, a particle oscillates between a periapsis $r_1$ and an apoapsis $r_2$, returning to its initial state within a finite amount of time. This section is dedicated to exploring the properties of periodic orbits around the charged bumblebee black hole.
	
\begin{figure*}[!htb]
		\centering
		
        \subfloat[$E=0.97$ and $Q=0.4$ ]{
        \includegraphics[width=0.45\textwidth]{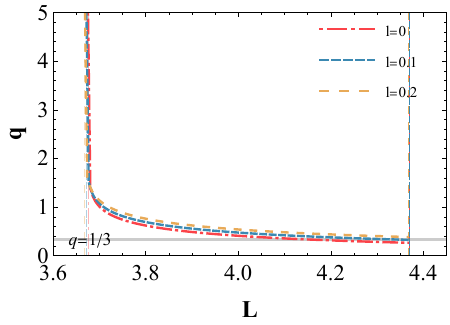} 
        \label{q_L_l}
        }
		\subfloat[$E=0.97$ and $l=0.1$ ]{
        \includegraphics[width=0.45\textwidth]{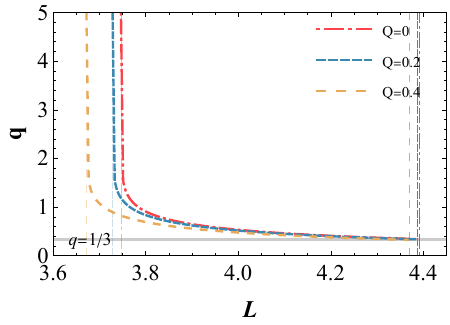} 
        \label{q_L_Q}
        }
		\caption{The rational frequency ratio $q$ as a function of the angular momentum $L$. In each panel, the vertical dashed lines of corresponding colors denote the allowed ranges of $L$ for bound motion.}
		\label{qofL}
	\end{figure*}

\begin{figure*}[!htb]
		\centering
		
        \subfloat[$L=3.6$ and $Q=0.2$]{
        \includegraphics[width=0.45\textwidth]{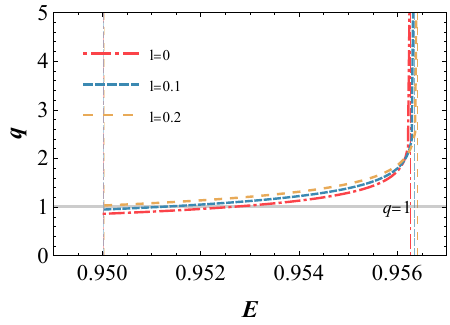} 
        \label{q_E_l}
        }
		\subfloat[$L=3.8$ and $l=0.1$]{
        \includegraphics[width=0.45\textwidth]{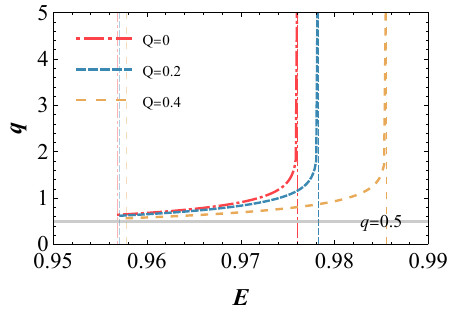} 
        \label{q_E_Q}
        }
		\caption{The rational frequency ratio $q$ as a function of the particle's energy $E$. In each panel, the vertical dashed lines of corresponding colors denote the allowed ranges of $E$ for bound motion.}
		\label{qofE}
	\end{figure*}

	For a particle moving in the equatorial plane, its dynamics are determined by the interplay between its radial and angular motions. Let $\Delta\phi$ denote the total azimuthal angle accumulated as the particle moves from one apoapsis to the next (i.e., completing one full radial period). This angle can be calculated by integrating the relationship between angular and radial motion:
    \begin{equation}\label{Deltaphi}
    \begin{split}
    \Delta\phi &= \oint d\phi = 2 \int_{\phi_1}^{\phi_2} d\phi \\
    &= 2 \int_{r_1}^{r_2} \frac{\dot{\phi}}{\dot{r}} dr = 2\sqrt{1+l} \int_{r_1}^{r_2} \frac{L}{r^2\sqrt{E^2 - V_{\text{eff}}^2}} dr.
    \end{split}
    \end{equation}
    The defining characteristic of a periodic orbit is that the ratio of the particle's radial oscillation frequency, $\omega_r$, to its angular rotation frequency, $\omega_\phi$, must be a rational number. Following common practice in the study of periodic orbits \cite{Levin:2008mq}, we introduce an ideal parameter $q$ to relate these frequencies to three integers $(w, z, v)$, as expressed by:
    \begin{align} \label{qFun}
		q = \frac{\omega_\phi}{\omega_r} - 1 = \frac{\Delta \phi}{2\pi} - 1 = w + \frac{v}{z} .
	\end{align}
    To avoid orbital degeneracy, $z$ and $v$ are coprime integers satisfying $1 \le v < z$. These integers provide a geometric interpretation of the orbit's trajectory:
    \begin{itemize}
    \item The Whirl number ($w$) quantifies the additional revolutions the particle makes around the central object near the periapsis during one radial period. A higher $w$ signifies a denser whirling behavior of the particle's trajectory in the vicinity of the central body.
    \item The Zoom number ($z$) corresponds to the number of ``petals'' or cusps in the complete periodic orbit. A larger value of $z$ indicates a more intricate and complex orbital trajectory.
    \item The Vertex number ($v$) describes the arrangement or spacing of the orbital petals. It determines how many petals the particle jumps when moving from one apoapsis to the next. For instance, if $z=4$ and $v=1$, the petals are arranged sequentially. If $v=3$, the petals are arranged with a gap.
    \end{itemize}
    The rational frequency ratio $q$ is a key observable that characterizes the degree of whirling in zoom-whirl orbits. By numerically solving Eq.~\eqref{qFun} for $q$, we can investigate how this parameter varies with the particle's energy $E$ and angular momentum $L$, as illustrated in Figs.~\ref{qofL} and \ref{qofE}. Specifically, $(w,z,v)=(n,1,1)$ is equivalent to $(w,z,v)=(n+1,1,0)$, both of which correspond to $q=n+1$.
    
    In Fig.~\ref{qofL}, with energy fixed at $E=0.97$, $q$ decreases monotonically as the angular momentum $L$ increases. Physically, a larger $L$ provides a stronger potential barrier, preventing the particle from approaching the event horizon too closely, thus reducing the whirl behavior. Conversely, as $L$ decreases toward its critical lower bound (represented by the vertical dashed lines), $q$ exhibits an ``explosive'' growth. This divergence occurs because the orbit approaches the separatrix (the boundary between bound and plunging orbits), where the radial period  tends to infinity while the particle spends an increasing amount of time whirling near the unstable circular orbit. The Lorentz-violating parameter $l$ and the charge $Q$ also significantly affect the frequency ratio $q$. As illustrated in Fig.~\ref{q_L_l}, the entire $q(L)$ curve shifts systematically to the upper-left as $l$ increases. This shift indicates that for a fixed $q$ (i.e., for the same orbital topology $(w, z, v)$), an increase in $l$ allows the orbit to exist at a lower angular momentum in the large-$q$ regime, whereas the opposite effect is observed in the small-$q$ regime. While in Fig.~\ref{q_L_Q}, an increase in $Q$ shifts the $q(L)$ curve systematically toward the left. This shift indicates that for a fixed $q$ (i.e., for the same orbital topology), a larger $Q$ enables the orbit to exist at a relatively lower angular momentum $L$.
    
    Fig.~\ref{qofE} explores the dependence of $q$ on the energy $E$. Here, $q$ is an increasing function of $E$, as higher energy allows the particle to overcome the potential barrier and penetrate deeper into the strong-field region near the horizon, thereby increasing the azimuthal precession. Similar to the behavior in Fig.~\ref{qofL}, $q$ diverges for the same reason as $E$ approaches a critical upper bound. As illustrated in Fig.~\ref{q_E_l}, the $q(E)$ curve shifts systematically toward the upper-right as $l$ increases. This shift indicates that for a fixed $q$ (i.e., for the same orbital topology), an increase in $l$ allows the orbit to exist at a lower energy $E$ in the small-$q$ regime, whereas the opposite effect is observed in the large-$q$ regime. While in Fig.~\ref{q_E_Q}, an increase in $Q$ shifts the $q(E)$ curve systematically toward the right. This shift indicates that for a fixed $q$ (i.e., for the same orbital topology), a larger $Q$ requires a higher energy $E$ for the orbit to exist.

\begin{figure*}[!htb]
		\centering
		
        \subfloat[$L=3.6$  and $Q=0$]{
        \includegraphics[width=0.45\textwidth]{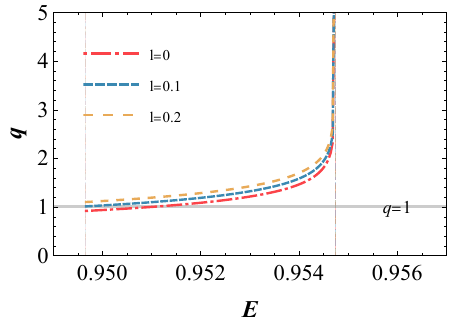} 
        \label{Q0_q_E_l}
        }
		\subfloat[$E=0.97$  and $Q=0$]{
        \includegraphics[width=0.45\textwidth]{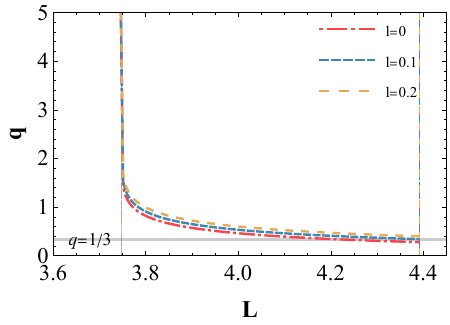} 
        \label{Q0_q_E_Q}
        }
		\caption{The rational frequency ratio $q$ as a function of orbital parameters for $Q=0$. In both panels, the vertical dashed lines represent the boundaries of the allowed ranges for bound orbits.}
		\label{Q0_qofE}
	\end{figure*}

    We examine the special case of vanishing charge ($Q=0$) in Fig.~\ref{Q0_qofE}. Unlike the charged cases shown in Fig.~\ref{qofL} and \ref{qofE}, the behavior of the frequency ratio $q$ in the $Q=0$ limit exhibits a distinct characteristic: the curves undergo a purely vertical shift as $l$ varies, without any horizontal displacement of the orbital parameter boundaries. As discussed in Sec.~\ref{Q0}, the effective potential $V_{\text{eff}}$ becomes independent of $l$ in this limit, meaning the allowed ranges for bound orbits (denoted by the fixed vertical dashed lines in Fig.~\ref{Q0_qofE}) and the radial turning points $r_1$ and $r_2$ for a given $(E, L)$ remain identical to those in the Schwarzschild spacetime. However, the azimuthal precession $\Delta \phi$ is explicitly modulated by the factor $\sqrt{1+l}$ according to Eq.~\eqref{Deltaphi}. Consequently, for a fixed energy $E$ and an angular momentum $L$, an increase in $l$ leads to a systematic increase in the total accumulated angle $\Delta \phi$ per radial period, thereby shifting the rational frequency ratio $q$ upward. This vertical scaling of $q$ provides a clean signature of bumblebee gravity, demonstrating that Lorentz violation can significantly alter the orbital topology and the resulting gravitational waveform even when the effective potential remains unchanged.

\begin{figure*}[!htb]
		\centering
		\subfloat{
			\begin{minipage}[b]{1\textwidth}
				\includegraphics[width=0.325\textwidth]{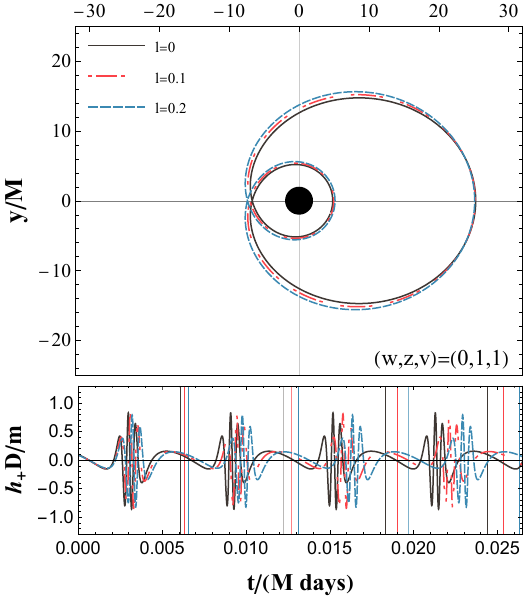} 
				\includegraphics[width=0.325\textwidth]{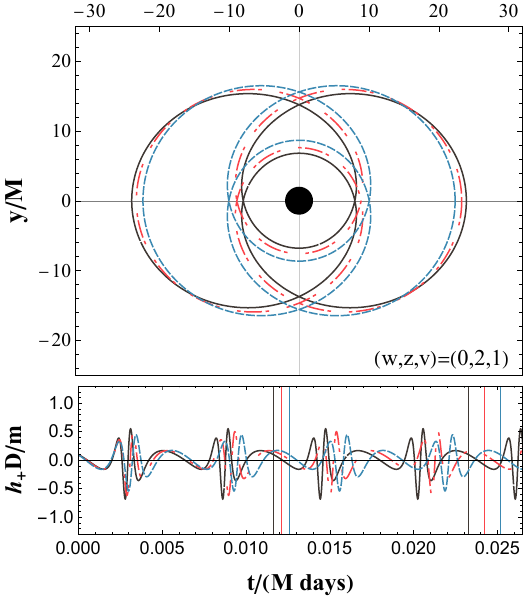} 
				\includegraphics[width=0.325\textwidth]{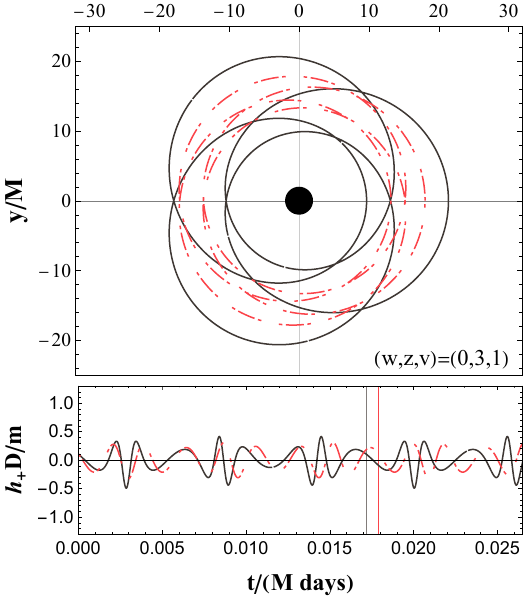} 
				\\
				\includegraphics[width=0.325\textwidth]{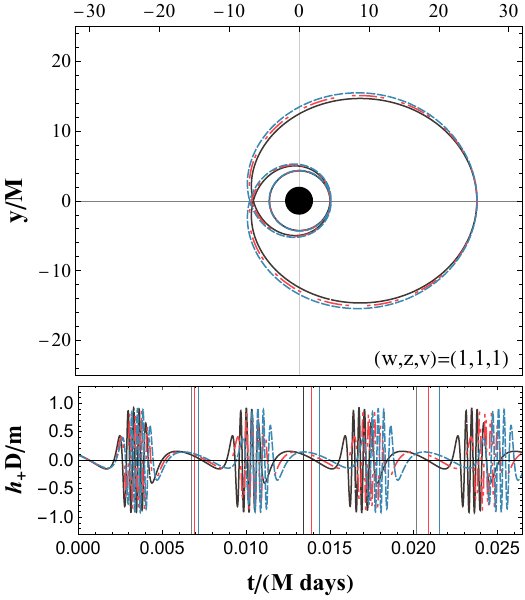} 
				\includegraphics[width=0.325\textwidth]{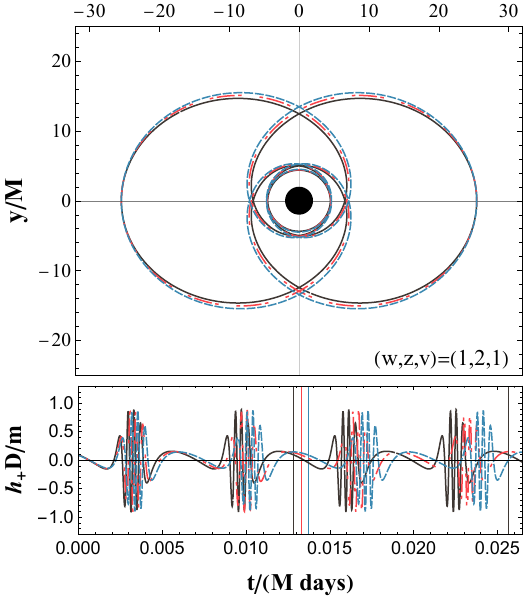} 
				\includegraphics[width=0.325\textwidth]{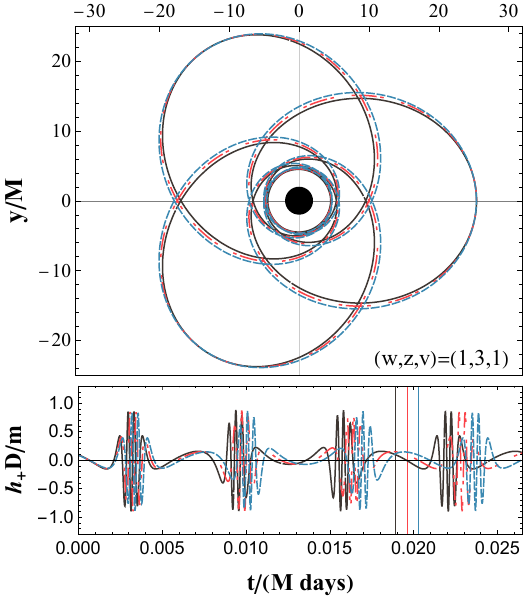} 
                \\
				\includegraphics[width=0.325\textwidth]{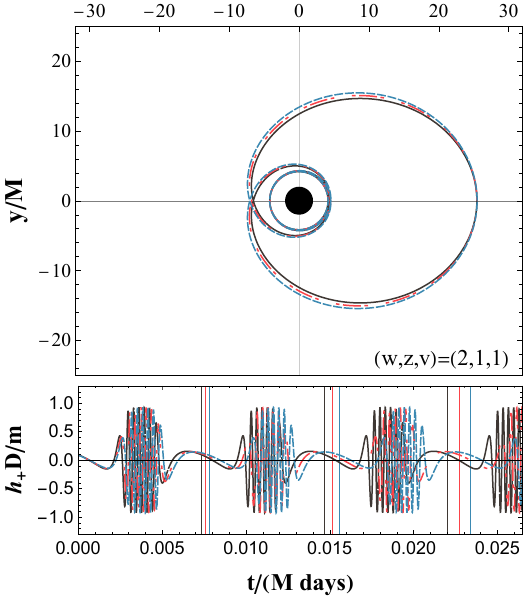} 
				\includegraphics[width=0.325\textwidth]{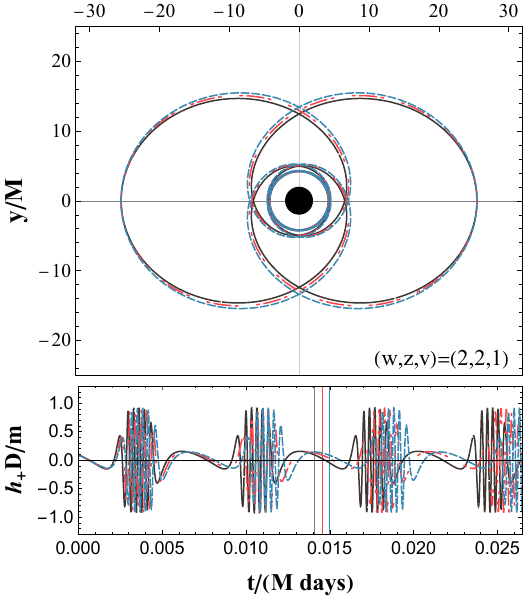} 
				\includegraphics[width=0.325\textwidth]{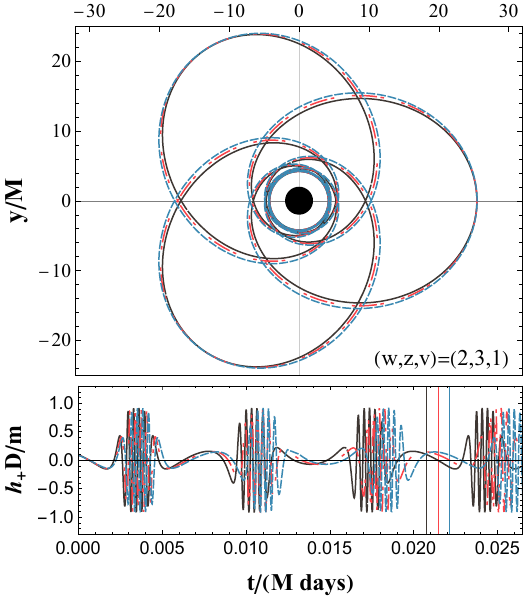} 
			\end{minipage}
		}
		\caption{Periodic orbits (upper panels) and corresponding gravitational waveforms (lower panels) in the charged bumblebee black hole with charge $Q = 0.4$ and energy $E = 0.97$, classified by the taxonomy $(w, z, v)$. In the waveform plots, vertical lines are used to separate each full orbital cycle.}
		\label{POGW_l}
	\end{figure*}

\begin{figure*}[!htb]
		\centering
		\subfloat{
			\begin{minipage}[b]{1\textwidth}
				\includegraphics[width=0.325\textwidth]{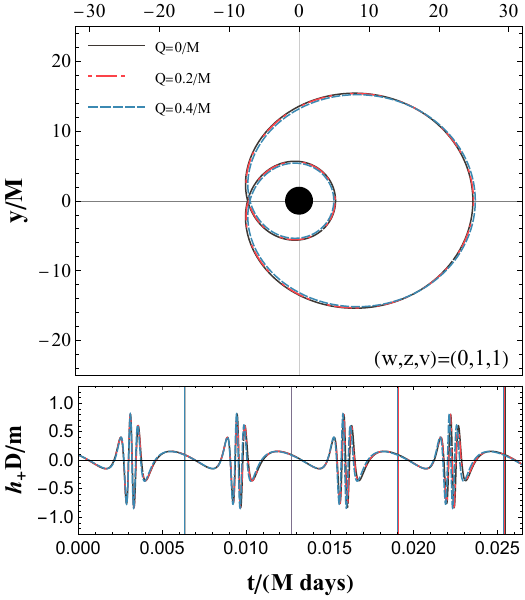} 
				\includegraphics[width=0.325\textwidth]{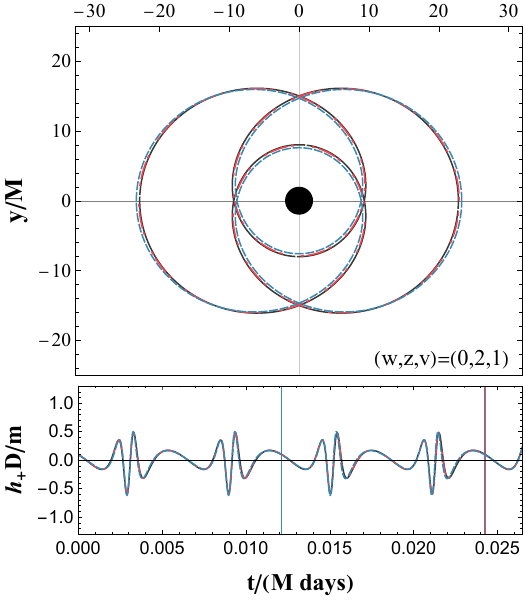} 
				\includegraphics[width=0.325\textwidth]{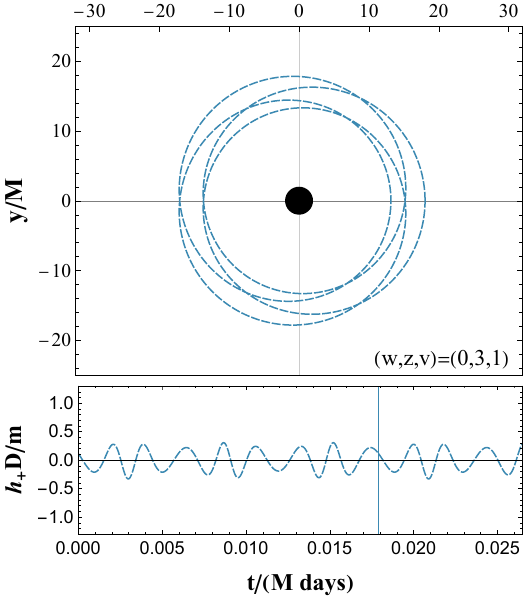} 
				\\
				\includegraphics[width=0.325\textwidth]{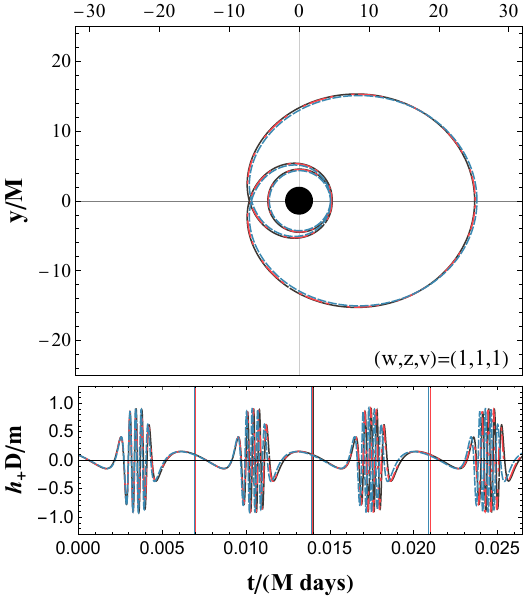} 
				\includegraphics[width=0.325\textwidth]{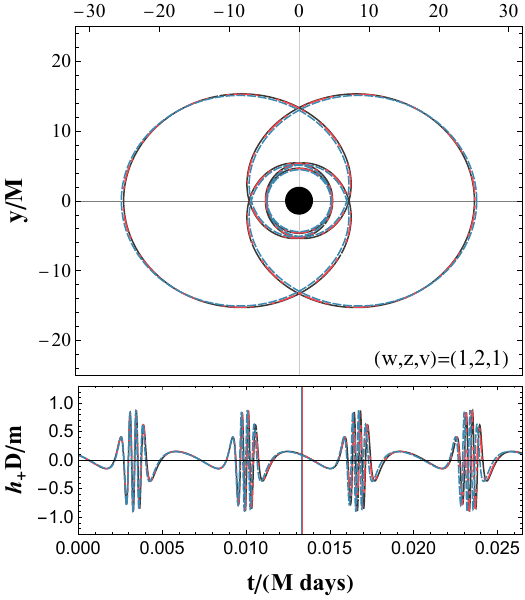} 
				\includegraphics[width=0.325\textwidth]{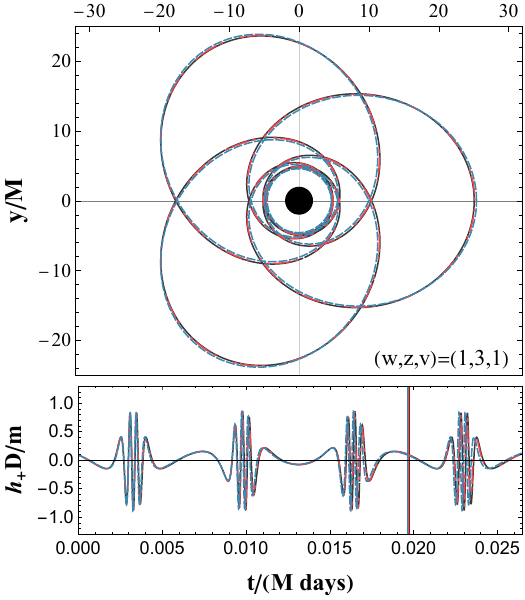} 
                \\
				\includegraphics[width=0.325\textwidth]{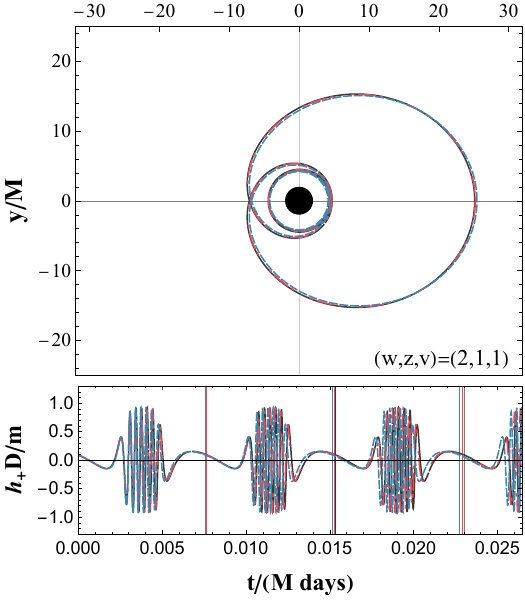} 
				\includegraphics[width=0.325\textwidth]{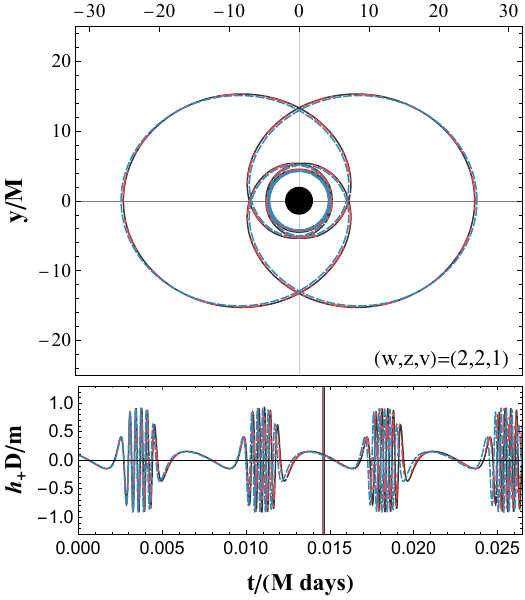} 
				\includegraphics[width=0.325\textwidth]{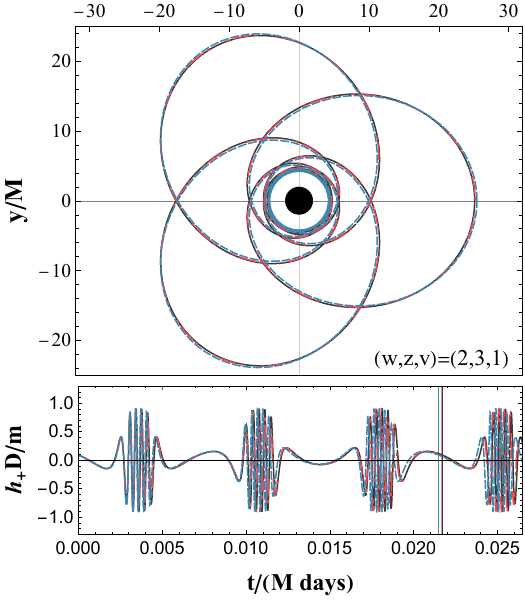} 
			\end{minipage}
		}
		\caption{Periodic orbits (upper panels) and corresponding gravitational waveforms (lower panels) in the charged bumblebee black hole with Lorentz-violating parameter $l = 0.1$ and energy $E = 0.97$, classified by the taxonomy $(w, z, v)$. In the waveform plots, vertical lines are used to separate each full orbital cycle.}
		\label{POGW_Q}
	\end{figure*}

	To further visualize the influence of Lorentz-violating parameter $l$, charge $Q$ and the rational ratio $q$ on the orbital geometry, we present a selection of periodic orbits in Figs.~\ref{POGW_l} and \ref{POGW_Q}.
    It is worth noting that, in order to maintain consistency with the gravitational wave section, we use non-normalized parameters when plotting the figures.
    As shown in the trajectory plots (the upper panels of each sub-figure), the zoom number $z$ dictates the number of radial petals. For instance, from left to right in the first row, we observe the transition from a single-petal orbit $(0,1,1)$ to a three-petal structure $(0,3,1)$. The whirl number $w$ (down each column) represents the number of additional revolutions performed near the periapsis. In $(1,1,1)$, a single inner loop is visible, while $(2,1,1)$ exhibits two complete inner whirls before the particle zooms out. The periodic orbits are further influenced by the Lorentz-violating parameter and black hole charge. As illustrated in Figs.~\ref{POGW_l} and \ref{POGW_Q}, varying these parameters leads to discernible shifts in the trajectories, such as modifications to the radial extent and the precise locations of the periapsis and apoapsis. These shifts encapsulate the direct response of the periodic motion to the modified spacetime geometry.

    We remark that some orbits corresponding to certain parameters are missing in the $(0,3,1)$ panels of Figs.~\ref{POGW_l} and \ref{POGW_Q}.
    This arises because the minimum value of $q$ becomes larger than $1/3$ as $Q$ or $l$ varies, as can be seen in Figs.~\ref{q_L_l} and \ref{q_L_Q}, which consequently leads to the absence of the $(0,3,1)$ orbit.

	\section{Gravitational Waveforms from Periodic Orbits}\label{section5}
    While the orbital dynamics discussed in the previous sections provide the theoretical foundation, the gravitational waveforms constitute the actual observables for detectors. To bridge the gap between theoretical trajectories and potential observations, this section is dedicated to outlining the framework and methodology for calculating the gravitational-wave signals emitted from these periodic orbits. It should be noted that in this section, unlike the previous ones, the parameters we use are unnormalized.
    
    The computation of gravitational waveforms for EMRIs is typically based on a key premise: the orbital parameters evolve on a much longer timescale than the orbital period. Owing to this condition, the energy and angular momentum of the secondary object can be treated as effectively conserved quantities over several orbital revolutions. Consequently, its motion can be accurately described as a geodesic in the background spacetime over a relatively short time window. The computational framework presented in this work is built upon this adiabatic approximation \cite{Hughes:1999bq}, which, by neglecting the back-reaction of gravitational radiation, allows for the independent solution of the periodic orbit (as was established in the preceding sections). This solution subsequently serves as the basis for waveform derivation.

    To arrive at the gravitational waveforms\cite{Babak:2006uv}, we apply the quadrupole formula directly to the trajectories derived in the previous section. This method effectively translates the numerical solutions of the geodesic equations into observable gravitational-wave signals.
    
    In the weak-field regime, the spacetime metric $g_{\mu\nu}$ can be decomposed as a small perturbation $h_{\mu\nu}$ propagating on a flat Minkowski background $\eta_{\mu\nu}$, such that $g_{\mu\nu} = \eta_{\mu\nu} + h_{\mu\nu}$. Under this approximation, the gravitational radiation is described by the spatial components of the metric perturbation $h_{ij}$, which is sourced by the symmetric and trace-free (STF) mass quadrupole moment $I_{ij}$. The fundamental quadrupole formula relates these quantities as follows \cite{Babak:2006uv}:
    \begin{equation}\label{hfromI}
    h_{ij} = \frac{2}{D_L} \frac{d^2 I_{ij}}{dt^2},
    \end{equation}
    where $D_L$ denotes the luminosity distance from the EMRI system to the detector. For a compact object of mass $m$ moving along a trajectory $Z^i(t)$, the quadrupole moment is explicitly given by \cite{Thorne:1980ru}:
    \begin{equation}
    I^{ij} = \int x^i x^j T^{tt}(t, x^i) d^3x,
    \end{equation}
    where
    \begin{equation}
    T^{tt}(t, x^i) = m \delta^{3}(x^i - Z^i(t))
    \end{equation}
    is the $tt$-component of the stress-energy tensor for the small object. The process of translating the orbital motion into gravitational waveforms requires careful consideration of coordinate systems. The orbital dynamics, initially solved in $(r, \theta, \phi)$, are first mapped onto a standard Cartesian coordinate system $(x, y, z)$ using the following relations:
    \begin{equation}
    x = r \sin\theta \cos\phi, \,  y = r \sin\theta \sin\phi, \, z = r \cos\theta.
    \end{equation}
    Consequently, the metric perturbations $h_{ij}$ can be derived from Eq. \eqref{hfromI} as
    \begin{equation}
    h_{ij} = \frac{2m}{D_L} (a_i x_j + a_j x_i + 2v_i v_j), 
    \end{equation}
    where $a_i$ and $v_i$ correspond to the acceleration and velocity components of the object, respectively. Then, we need to introduce another coordinate system $(X, Y, Z)$ that is detector-adapted, with its origin coinciding with the $(x, y, z)$ system at the black hole center \cite{Poisson:2014}. The basis vectors of $(X, Y, Z)$, expressed in terms of the original coordinate system $(x, y, z)$, are:
    \begin{align}
    e_X &= (\cos\zeta, -\sin\zeta, 0),\\
    e_Y &= (\sin\iota \sin\zeta, -\cos\iota \cos\zeta, -\sin\iota),\\
    e_Z &= (\sin\iota \sin\zeta, -\sin\iota \cos\zeta, \cos\iota),
    \end{align}
    where $\iota$ and $\zeta$ correspond to the orbital inclination with respect to the $X-Y$ plane and the longitude of the periapsis in the orbital plane, respectively \cite{Barack:2003fp}. From the full metric perturbation $h_{ij}$, two independent polarization modes, $h_+$ (plus polarization) and $h_\times$ (cross polarization), are extracted. These modes characterize the observable strains measurable by an interferometer. They are obtained by projecting $h_{ij}$ onto the detector frame:
    \begin{align}
    h_+ &= \frac{1}{2} (e_X^i e_X^j - e_Y^i e_Y^j) h_{ij},\\
    h_\times &= \frac{1}{2} (e_X^i e_Y^j + e_Y^i e_X^j) h_{ij}.
    \end{align}
    Alternatively, these polarizations can be formulated using intermediate components $h_{\zeta\zeta}$, $h_{\iota\iota}$, and $h_{\iota\zeta}$ \cite{Babak:2006uv}, which are defined in the detector frame as combinations of the $h_{ij}$ components as
    \begin{widetext}
    \begin{align}
    h_{\zeta\zeta} &= h_{xx} \cos^2\zeta - h_{xy} \sin 2\zeta + h_{yy} \sin^2\zeta,\\
    h_{\iota\iota} &= \cos^2\iota (h_{xx} \sin^2\zeta + h_{xy} \sin 2\zeta + h_{yy} \cos^2\zeta) + h_{zz} \sin^2\iota - \sin 2\iota (h_{xz} \sin\zeta + h_{yz} \cos\zeta),\\
    h_{\iota\zeta} &= \frac{1}{2} \cos\iota (h_{xx} \sin 2\zeta + 2h_{xy} \cos 2\zeta - h_{yy} \sin 2\zeta) + \sin\iota (h_{yz} \sin\zeta - h_{xz} \cos\zeta).
    \end{align}
    \end{widetext}
    In terms of these components, the two independent polarization states are then expressed as
    \begin{align}
    h_+ &= \frac{1}{2} (h_{\zeta\zeta} - h_{\iota\iota}),\\
    h_\times &= h_{\iota\zeta}.
    \end{align}
    
    The bottom panels of each sub-figure in Figs.~\ref{POGW_l} and \ref{POGW_Q} present the corresponding gravitational waveforms extracted from these periodic orbits. Several key features can be observed from the numerical results.
    
    Firstly, the complexity of the waveform is directly governed by the orbital taxonomy $(w, z, v)$. For the simplest $(0, 1, 1)$ orbit, the waveform exhibits a relatively simple, repeating structure. However, as the zoom number $z$ increases (moving from left to right in the top row), the waveform develops more sub-structures within each full orbital cycle, reflecting the multiple petals in the trajectory. More strikingly, for orbits with a non-zero whirl number $w$ (e.g., the $(1, 1, 1)$ and $(2, 1, 1)$ cases), the waveforms exhibit high-frequency bursts or whirl-phase oscillations near the periapsis. During these phases, the particle orbits rapidly near the black hole, leading to a significant increase in the gravitational-wave frequency and amplitude, which is a characteristic signature of zoom-whirl dynamics.

\begin{figure}[!htb]
		\includegraphics[width=0.325\textwidth]{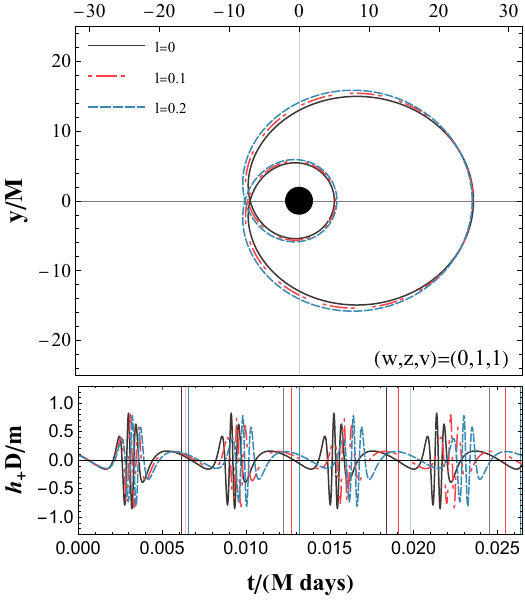} 
		\caption{Periodic orbit (upper panel) and corresponding gravitational waveform (lower panel) for the $(w, z, v)=(0, 1, 1)$ orbit with $Q=0$ and $E=0.97$. Vertical lines in the waveform plot separate each full orbital cycle.}
		\label{Q0_POGW_l}
	\end{figure}

    Secondly, the Lorentz-violating parameter $l$ exerts a significant influence on the phase evolution of the gravitational waves. Comparing the curves for $l=0, 0.1$, and $0.2$ in Fig.~\ref{POGW_l}, we observe a clear ``dephasing'' effect. As $l$ increases, the peak positions of the waveforms shift progressively to the right (positive time direction), indicating an increase in the orbital period. This cumulative dephasing is a crucial observational signature; even a small Lorentz-violating parameter can lead to a large phase mismatch over a long observation period, making it a potentially detectable signature for future space-based gravitational-wave interferometers. This dephasing effect remains qualitatively identical in the vanishing charge limit ($Q=0$), as illustrated in Fig.~\ref{Q0_POGW_l}.
    
    Thirdly, the effect of the black hole charge $Q$ on the waveforms, as illustrated in Fig.~\ref{POGW_Q}, presents an intriguing contrast to that of the Lorentz-violating parameter. Unlike $l$, an increase in $Q$ tends to shift the waveform peaks in the opposite direction, effectively shortening the orbital period. This leads to a critical physical implication: the presence of a residual charge $Q$ could potentially offset or mask the dephasing signatures induced by Lorentz violation. Therefore, if we aim to use gravitational wave morphology to search for signals of bumblebee gravity, the charge $Q$ must be treated as a confounding factor. A highly precise analysis of the waveform's fine structure is required, as relying solely on the orbital period might lead to an underestimation of the true Lorentz-violating strength.

	\section{Conclusion and Discussion}\label{section6}
	
	In this study, we have systematically explored the periodic orbits and the resulting gravitational waveforms of a massive test particle orbiting a charged black hole in bumblebee gravity. This spacetime provides a well-motivated testing ground for investigating the astrophysical implications of spontaneous Lorentz symmetry breaking, parameterized by $l$, in the presence of an electric charge $Q$. 
    
    We began by analyzing the geodesic equations and the effective potential of a test particle in the equatorial plane. Our results indicate that both the Lorentz-violating parameter $l$ and the electric charge $Q$ exert an uplifting effect on the effective potential. Consequently, they expand the parameter space (energy $E$ and angular momentum $L$) that permits bound orbits, shifting the ISCO towards lower energy and angular momentum values. This demonstrates that the modified spacetime geometry enhances the gravitational trapping effect around the black hole. Furthermore, we conducted a detailed investigation of periodic orbits, which occur when the ratio of the radial to angular oscillation frequencies is a rational number. By classifying these orbits using the whirl, zoom, and vertex taxonomy $(w, z, v)$, we mapped out the dependence of the rational frequency ratio $q$ on the orbital parameters and the background spacetime parameters. 
    
    Building upon this analysis, a particularly intriguing finding emerges in the $Q=0$ limit. In this uncharged scenario, the radial effective potential reduces exactly to that of a Schwarzschild black hole. Consequently, the standard characteristics of the ISCO—such as its radius, energy, and angular momentum—are completely degenerate with those in general relativity, making the Lorentz-violating parameter $l$ entirely masked. However, our investigation demonstrates that periodic orbits successfully break this degeneracy. Even when the potential landscape is indistinguishable from Schwarzschild, the azimuthal precession is explicitly modified by a factor of $\sqrt{1+l}$. This intrinsic difference forces the rational frequency ratio $q$ to deviate from the Schwarzschild baseline, manifesting as a distinct vertical scaling. The vertical scaling of $q$ provides a clean signature of bumblebee gravity, demonstrating that Lorentz violation can significantly alter the orbital topology and the resulting gravitational waveform. Therefore, periodic orbits serve not merely as a dynamical curiosity, but as a unique and powerful diagnostic tool capable of revealing spacetime deviations that remain otherwise completely obscured in standard effective potential analyses. 
    
    Finally, we bridged these theoretical trajectories to observable signals by computing the gravitational waveforms using the quadrupole formula. The waveforms exhibit complex substructures, with high-frequency bursts corresponding to the periapsis passages in orbits with non-zero whirl numbers. Crucially, we identified significant phase evolution effects induced by the spacetime parameters. An increase in the Lorentz-violating parameter $l$ shifts the waveform peaks to the right (lengthening the orbital period), whereas an increase in the charge $Q$ shifts them to the left. 
    
    This contrasting effect on the gravitational-wave phase has important observational implications. It suggests that a residual black hole charge could partially mask or offset the dephasing signatures induced by Lorentz violation, potentially leading to an underestimation of the true Lorentz-violating strength. Therefore, in the context of EMRIs detectable by future space-based observatories like LISA, Taiji, or TianQin, a highly precise, multi-parameter template matching will be essential. Relying solely on the cumulative phase shift might lead to parameter degeneracies. Future work could extend this analysis by considering rotating black hole solutions within bumblebee gravity, incorporating the radiation reaction for a full inspiral evolution, and performing rigorous parameter estimation studies to quantify the detectability of Lorentz violation in the presence of other astrophysical uncertainties.
	
	\begin{acknowledgements}
		This work is supported by National Natural Science Foundation of China (NSFC) with Grants No.12275087. 
	\end{acknowledgements}	
	\appendix

	\par
	\bibliographystyle{unsrt}

\end{document}